\documentclass[twocolumn,showpacs,prb,superscriptaddress,aps,floatfix]{revtex4-1}

\usepackage{rotating}
\usepackage{amsmath}
\usepackage{color}
\usepackage{graphicx}
\usepackage[active]{srcltx}
\usepackage[sort&compress]{natbib}

\newcommand{\be}{\begin{equation}}
\newcommand{\ee}{\end{equation}}
\newcommand{\bea}{\begin{eqnarray}}
\newcommand{\eea}{\end{eqnarray}}

\newcommand{\lille}{Institute for Electronics, Microelectronics, and
Nanotechnology (IEMN), CNRS-UMR 8520, Dept. ISEN, B.P. 60069, 59652
Villeneuve d'Ascq Cedex, France} 
\newcommand{\grenoble}{Institut Neel,
CNRS/UJF, 25 rue des Martyrs BP 166, B\^{a}timent D 38042 Grenoble
cedex 9 France} 
\newcommand{\sanseb}{Nano-Bio Spectroscopy Group and
ETSF Scientific Development Centre, Dpto. F\'isica de Materiales,
Universidad del Pa\'is Vasco, Centro de F\'isica de Materiales
CSIC-UPV/EHU-MPC and DIPC, Av. Tolosa 72, E-20018 San Sebasti\'an,
Spain} 
\newcommand{\fritzhaber}{Fritz-Haber-Institut der Max-Planck-Gesellschaft, 
Berlin, Germany}
\newcommand{\erlangen}{Lehrstuhl f\"ur Theoretische
Festk\"orperphysik, Universit\"at Erlangen-N\"urnberg, Staudtstrasse 7
B2, D-91058 Erlangen, Germany}
\newcommand{\rome}{Dipartimento di Fisica, Universit\`a di Roma ``Tor
Vergata'', via della Ricerca Scientifica 1, I-00133 Roma, Italy}
\newcommand{\ikerb}{Ikerbasque, Basque Foundation for Science, E-48011 Bilbao, Spain}

\begin{document}
\title{
Coupling of excitons and defect states in boron-nitride nanostructures 
}

\author{C. Attaccalite}
\affiliation{\grenoble}

\author{M. Bockstedte} 
\affiliation{\erlangen}
\affiliation{\sanseb}

\author{A. Marini}
\affiliation{\rome}
\affiliation{\ikerb}
\affiliation{\sanseb}

\author{A. Rubio} 
\affiliation{\sanseb}
\affiliation{\fritzhaber}

\author{L. Wirtz}
\affiliation{\lille}

\begin{abstract}
The signature of defects in the optical spectra of hexagonal boron
nitride (BN) is investigated using many body perturbation theory. A
single BN-sheet serves as a model for different layered
BN-nanostructures and crystals. In the sheet we embed prototypical
defects such as a substitutional impurity, isolated Boron and Nitrogen vacancies, and the di-vacancy. Transitions between the deep defect levels
and extended states produce
characteristic excitation bands that should be responsible for the emission
band around $4$~eV, observed in luminescence experiments. In addition,
defect bound excitons occur that are consistently treated in our ab
initio approach along with the ``free'' exciton. For defects in strong 
concentration, the co-existence of both bound and free excitons adds 
sub-structure to the main exciton peak and provides an explanation for 
the corresponding feature in cathodo and photo-luminescence spectra.

\end{abstract}           


\maketitle

\section{Introduction}
Boron nitride crystals and nano structures are excellent wide band gap
materials for the realization of UV-lasers.\cite{patent} Defects
formed during the BN-synthesis or introduced as impurities can bind
excitons and act as centers for their recombination. This leads
to (desired or undesired) luminescence bands and losses in the photon 
frequency range of interest. In fact, recent
experiments~\cite{watanabe1,watanabe2,watanabe3,PhysRevB.75.085205,Museur2008,PhysRevB.78.155204,loiseau,PhysRevB.77.235422} have given evidence
that defects strongly effect the luminescence of 
BN-based materials.  In order to achieve optimal optical properties, the
relevant defects have to be identified such that methods can be devised
to tune their concentration in the material.

Boron nitride (BN) is known to exist in three different crystals
structures, namely hexagonal BN (h-BN), cubic BN (c-BN), and wurtzite
BN (w-BN). Among these structures, h-BN is the stable one at room
temperature and ambient pressure. Like graphite, it
consists of stacked BN-layers with a honeycomb structure. 
BN nanostructures\cite{xblasereview,wirtz_review} ranging from
single and multi-wall nanotubes to BN-fullerens and nanocones can be envisaged
as having been tailored from a single BN sheet. Unlike in graphite,
the partially ionic character of the BN bond results in a wide band-gap
of about 6.5 eV for bulk h-BN\cite{Arnaud2006}. The combination of a large gap with
a strong electron-hole attraction makes the optical properties of BN
nanostructures largely independent of the layer
arrangement.\cite{PhysRevLett.96.126104,wirtz_review} 
A single layer of h-BN is thus
a generic system for studying the optical properties of
crystalline and nanostructured BN including the exciton-defect interaction.

Detailed experiments on the cathodo and photo-luminescence of h-BN crystals
and BN nanotubes have been performed in the recent years
\cite{watanabe1,watanabe2,watanabe3,PhysRevB.75.085205,Museur2008,PhysRevB.78.155204,loiseau,PhysRevB.77.235422,zhi:213110}.
The origin of some of the features in the BN luminescence spectra 
remains unsettled. For instance, the occurrence of a defect-related emission
band at $4$~eV, composed of regularly spaced peaks compatible with phonon
replicas\cite{PhysRevB.75.085205,PhysRevB.78.155204,jaffrennou:116102}, can be associated with 
a deep level impurity. However, the defect that is responsible for this
deep level has not yet been identified. Furthermore, the emission band 
at $5.77$~eV shows different sub-peaks on the low energy side. 
The origin of its sub-structure is unclear, since the main peak 
can be understood as being due to the recombination of a Frenkel-type
exciton\cite{Arnaud2006} with a degenerate recombination
line.\cite{PhysRevLett.96.126104,PhysRevLett.100.189701}
Interpretations of the experimental findings are based on different
mechanisms: quasi-donor-acceptor pairs\cite{Museur2008}, excitons
coupled with structural defects
\cite{PhysRevB.77.235422,jaffrennou:116102}, breaking of the exact
hexagonal symmetry (e.g., by the presence of defects)
\cite{PhysRevLett.100.189701}, or a dynamical Jahn-Teller
effect\cite{PhysRevB.79.193104} (symmetry breaking due to geometry relaxation
in the excited state).

Theoretical works on defects in BN-structures have so far focused on
the electronic and structural properties of point defects
\cite{azevedo,PhysRevB.67.113407,PhysRevB.63.125205,hou:076103,PhysRevB.81.075125,PhysRevB.75.193409,PhysRevB.76.014405}.
The abundance (related to the formation energy) and stability of different 
defects has 
been addressed with detailed density-functional theory (DFT)
calculations.\cite{PhysRevB.63.125205,PhysRevB.80.161404,PhysRevB.66.014112,PhysRevB.81.153407,PhysRevB.75.094104,zobelli}.
The position of deep levels within the fundamental band gap has been
calculated on the DFT level.  However,
DFT calculation suffer from the well-known underestimation of the
band-gap. Therefore, besides the neglect of the strong excitonic
effects in BN, (de)excitation spectra based on DFT energy levels may 
not be reliable enough for comparison with experimental spectra.

In the present paper we address the optical properties of 
point defects in BN going beyond the approximation of DFT.  
We focus on a substitutional
carbon impurity on the nitrogen site, the isolated vacancies and the
BN di-vacancy that are common defects in BN.
We will demonstrate that defects are responsible for the emission
lines around $4$~eV. Their presence can also explain the
experimentally observed splitting of the main exciton peak.
The paper is organized as
follows: in section \ref{computational} we summarize the computation
methods employed in this work; in section \ref{dft} we present the
ground state results and the quasi-particle(QP) band structures; in
section \ref{optical} we investigate the role of defects on the optical
properties and finally in section \ref{conclusion} we 
compare our results with experiments.
\section{Computational Methods}
\label{computational}
\begin{figure*}[t]
 \centering
 \includegraphics[width=0.99\textwidth]{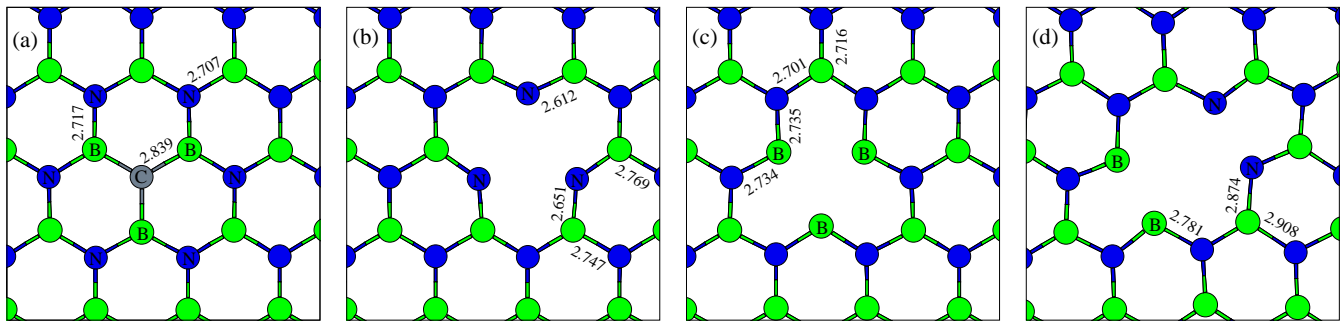}
\caption{(color online) Geometry of the defects in  BN mono-layer with defects:
 $(a)$ $C_N$; $(b)$ $V_B$; $(c)$ $V_N$; $(d)$ $V_{BN}$. Defect induced
changes of BN bond distances are indicated (in atomic units) and have to
be compared with the BN distance of the perfect sheet ($2.724 a.u.$).
\label{defects_structure}}
\end{figure*}
Density functional theory (DFT) within the local spin density (LSDA) or gradient
corrected approximations (GGA) for the exchange-correlation energy has been
widely used to study defects in semiconductors and insulators. Although 
DFT is in principle exact, it is limited to the ground state and as a closed
expression for the exchange-correlation energy functional is unknown,
approximations like the LSDA or GGA have to be employed.  Albeit theoretically
well-founded, these approximate functionals bear inherent limitations like the
well-known band gap problem which concomitantly affect the prediction of
defect levels, in particular in wide band gap insulators.  In the last years,
many-body perturbation theory within the GW approximation\cite{Aulbur19991}
has become a valid and accurate alternative to predict defects properties in
semiconductors and insulators
\cite{PhysRevLett.102.026402,PhysRevB.81.113201,PhysRevB.51.7464} starting
from the DFT ground state. Moreover, in going beyond the ground state, the
optical properties of defects and the host insulator can be addressed by 
including the electron-hole interaction in the excited state via the 
Bethe-Salpeter equation (BSE) \cite{strinati2,PhysRevB.77.115118,bockstedte}.  

Here we use this approach to study the role of prototypical defects in 
the optical properties of BN nanostructures: a substitional carbon impurity on
the nitrogen sublattice ($\text{C}_\text{N}$), the boron vacancy
($\text{V}_\text{B}$), the nitrogen vacancy ($\text{V}_{\text{N}}$) and the
boron-nitrogen di-vacancy ($\text{V}_{\text{BN}}$).  Calculations proceed in
three steps: first we obtain the DFT ground state employing the LSDA
approximation and a plane wave pseudopotential method,\cite{pwscf} (see
\onlinecite{DFT} for details). Second we evaluate the quasi-particle band
structure within the G$_{0}$W$_{0}$\cite{Aulbur19991,RevModPhys.74.601,hybert} approach and third we address the optical
spectra via the BSE.\cite{strinati}
A very large supercell\cite{DFT} is employed to accommodate the defect and to
enable at the same time the description of defect bound and free
excitons. This becomes possible due to the strong localization of the bulk
exciton state in h-BN,\cite{Arnaud2006} in the layer, and tubes
\cite{PhysRevLett.96.126104,park2006}.

As it has been shown that a large class of
 impurities, when introduced into h-BN, is in a
 spin-polarized state and acquires a local magnetic moment\cite{PhysRevB.76.014405}, 
 we have performed spin-polarized calculations for the $\text{C}_\text{N}$, 
 $\text{V}_\text{B}$, and $\text{V}_\text{N}$ cases while
 this was not necessary for the di-vacancy $\text{V}_{\text{BN}}$  that does not
 display a net spin polarization.
The quasi-particle properties have been calculated starting from the Kohn-Sham
Hamiltonian and orbitals within the G$_0$W$_0$ approximation\cite{GW}: \be
E_{nk} = \epsilon_{nk} + Z_{nk} Re\Delta \Sigma_{nk}(\epsilon_{nk}),
\label{per}
\ee where $\Delta \Sigma = \Sigma-V^{\text{xc}}$ and $V^{\text{xc}}$ is the LSDA
exchange-correlation functional, $\epsilon_{nk}$ are the Kohn-Sham
eigenvalues, and $Z_{nk} = [1-\partial Re \Delta \Sigma / \partial w]^{-1}$
is the renormalization factor\cite{Aulbur19991}. The screened
electron-electron interaction $W$ has been evaluated within the random-phase
approximation (RPA) in
terms of the dielectric function $\epsilon_{G,G'}(q,\omega)$ using a
plasmon-pole model.\cite{epsilon}  
Neutral excitations are investigated on the basis of the
quasiparticle spectrum by solving the Bethe-Salpeter equation (see
\onlinecite{BSE} for details). Excitonic effects
are analysed through a comparison of the absorption
spectra for independent particles 
with absorption spectra where electron-hole attraction is included on the 
level of the BSE.

As we are studying an isolated monolayer employing three-dimensional
periodic boundary conditions, the size of the quasi particle energies
and the energy of the electron-hole attraction will depend on the
inter-sheet distance. In Ref.~\onlinecite{PhysRevLett.96.126104} we had
shown that the value of the gap of the pure h-BN sheet is 0.6 eV higher for
an inter-sheet distance is 80 a.u.\ than for an  inter-sheet distance of 
20 a.u.\ (value which we
are using in the calculations presented here). At the same time
we had observed an increase of the exciton binding energy with increasing
inter-sheet distance that almost exactly cancels  the variation of the band-gap.
We expect a similar cancellation effect in the presence of defects 
but cannot exclude
a minor effect of the inter-sheet distance on the absolute position
of defect peaks.

\section{DFT-ground state and Quasi-particle states}
\label{dft}
In this section we address the electronic and geometric structure of
the four defects and then compare the DFT-electronic structure with
the quasiparticle states obtained from G$_0$W$_0$ calculations.

In Fig.~\ref{defects_structure} the equilibrium geometry of the
C$_\text{N}$ and the vacancies V$_\text{N}$,V$_\text{B}$ and
V$_\text{B\/N}$ is shown along with the BN bond distances in the
vicinity of the defect center. In the case of subsitutional impurity all
the first neighbor B atoms of $C_N$ undergo a small symmetry
conserving outward relaxation of 0.113 a.u. (4\% of the BN-bond
distance) with respect to the ideal lattice positions. The defect
retains its threefold symmetry. In the case of $V_B$ and $V_N$ the the
outward displacement of the first-neighbor atoms is larger. More
importantly, in agreement with Ref.~\onlinecite{PhysRevB.76.014405},
we find that a (pseudo) Jahn-Teller effect breaks the three-fold
symmetry slightly. In the di-vacancy $V_{BN}$ case, the three-fold
symmetry is already reduced to a mirror symmetry by the defect
configuration itself. We observed that the first N and B neighbors
decrease their mutual distance upon relaxation, as shown in 
Fig.~\ref{defects_structure}d.  The defect geometries found here
are in good agreement with those reported in
Ref.~\onlinecite{PhysRevB.63.125205} and \onlinecite{azevedo}.

\begin{figure*}[ht]
 \centering
 \includegraphics[width=0.95\textwidth]{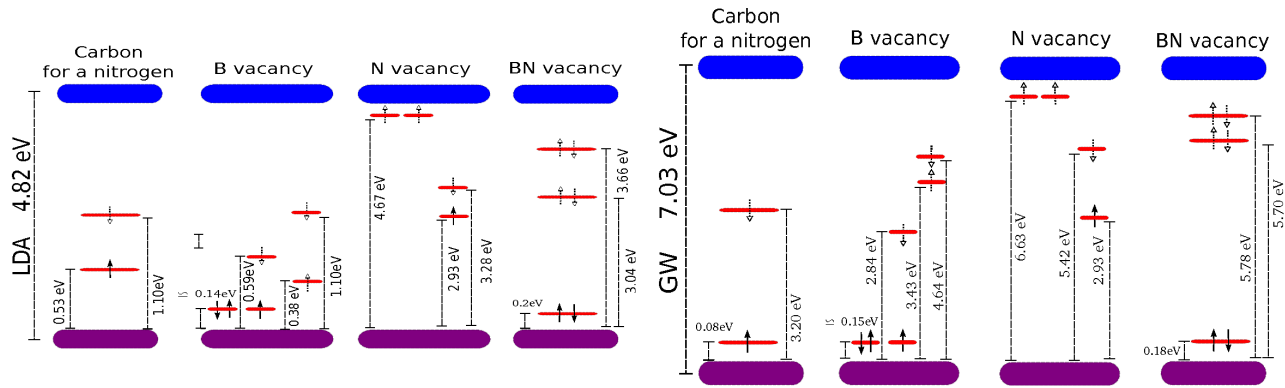}
 \caption{(color online) Quasiparticle energies of C$_\text{N}$,
 V$_\text{B}$, V$_\text{N}$, and V$_\text{B\/N}$ as obtained from LDA and 
 G$_0$W$_0$ calculations: schematic representation of energy position
 with in the band gap.  The filled arrows indicate levels occupied 
 with electrons with spin up or down. Dotted arrows indicate unoccupied states
 (holes) correspondingly.\label{ldagw_levels}}
\end{figure*}

As established already in earlier work, the C$_\text{N}$ and the
vacancies possess localized defect levels in the band gap. Here we
briefly review the electronic structure of the defects as obtained in
DFT-LSDA and then compare it with the quasiparticle states obtained on the
level of the G$_0$W$_0$ approximation where the 
DFT band gap-error is corrected.
Since the
prediction of defect emission and absorption spectra hinges on the
defect level positions, a correct prediction
of these quantities is a crucial point for the investigation of
optical properties. The defect levels as obtained within the
DFT-L(S)DA are reported in the left panel of Fig.~\ref{ldagw_levels}.

In the case of $C_N$ the defect levels arise from the three bonds of
the substitutional carbon impurity with its boron neighbors. These bonds
combine into two fully occupied states located within the valence band
and a third half-occupied localized state within the band gap. As a
result of the occupation with only one electron there is a
spin-splitting of  $~\sim 0.5$\,eV between the
occupied (spin-up) and unoccupied (spin-down) state.

The situation is more complex for the boron vacancy. In this case a
two-fold degenerate and a non-degenerate defect state are formed from the
unpaired orbitals at the three nitrogen neighbors. In agreement with
Ref.~\onlinecite{PhysRevB.66.014112,PhysRevB.63.125205,zobelli}, these
defect levels are located within the band gap. In the neutral vacancy the
two-fold degenerate level is occupied by three electrons and the
non-degenerate level remains empty. Due to the Jahn-Teller distortion
a small splitting of the degenerate
level arises in addition to the exchange-splitting of the half- and unoccupied
states. $V_\text{B}$ behaves as a triple acceptor. 

For $V_\text{N}$ we obtain a half-filled level in the band gap, and
two un-occupied levels almost degenerate in energy with the conduction
bands. Due to spin-splitting their counter parts of opposite spin are
found as resonances in the conduction band. Finally, in the case of the
di-vacancy V$_\text{BN}$, there are four orbitals originating from the
two boron and nitrogen neighbors. Only three are located in the band gap,
one being occupied by two electrons and the upper two remaining empty.

The above picture obtained from the Kohn-Sham eigenvalues gives a
qualitative description of the defect levels in the band gap. The
position of the states is, however, quantitatively affected by the
LSDA-band gap error and the approximate treatment of exchange with
this XC-functional as discussed above. Furthermore, the ordering
of occupied and unoccupied states on the Kohn-Sham level may be
different from that of the ``true'' quasi-particle energy levels
(see e.g.~\onlinecite{Aulbur19991}).
The G$_0$W$_0$ method yields true quasiparticle
energies and an improved description of exchange and screening
effects.  Due to the reduced dimensionality of the BN mono-layer there
is an incomplete screening of the Coulomb interaction, such that the
quasi-particle corrections are expected to be quite large.
We calculated the quasi-particle band structure for all the cases
considered above. The calculated band gap is in agreement
with the results of Ref.~\onlinecite{PhysRevLett.96.126104}. 
In the right panel of Fig.~\ref{ldagw_levels} the position of the defect quasi particle
energies within the band gap is shown schematically.  The G$_0$W$_0$
corrections to the LSDA levels depend on the orbital occupation and
on the wave-function character\cite{Aulbur19991}. 
The exchange splitting between occupied and unoccupied
defect levels is strongly enhanced and the ordering of defect levels 
within the band gap is altered in some cases. In all cases we find that 
occupied levels are pushed closer
to the valence bands and unoccupied ones closer to conduction
bands. For C$_\text{N}$ the spin-splitting is significantly increased
from $0.5$\,eV to $3.1$\,eV. A large exchange splitting is also
present in the case of V$_\text{N}$. For $V_\text{B}$ all un-occupied
levels are pushed up by more then $2$~eV, depending on the character
of the wave-function, and the level order is modified as well. In the
case of V$_\text{B\/N}$, un-occupied levels are found close to the
conduction band edge, while the occupied ones remain in their position
close to the valence band edge. This underlines the fact that the
di-vacancy contains both donor and acceptor levels. The general
picture obtained from G$_0$W$_0$ shows the clear deficiencies of the
DFT-LSDA in the description of defect level positions in the case of
BN, in particular, for unoccupied states. 
Vested with the defect quasiparticle spectrum we now turn to the
optical properties.

\section{Optical properties}
\label{optical}
\begin{figure}[ht]
 \centering
 \includegraphics[width=0.31\textwidth,angle=-90]{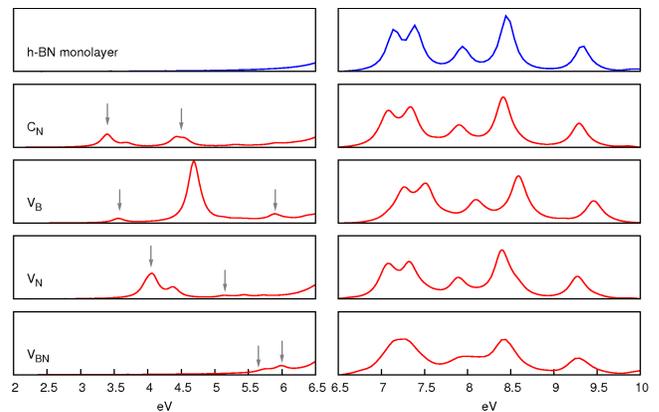}
 \caption{Independent-particle absorption spectra (arbitrary units)
based on the $G_0W_0$ quasi-particle band structure. In the 
left panel, the absorption cross section is increased by a factor 15
with respect to the right panel in order to make the defect-related 
peaks visible.
\label{optics_ip}.}
\end{figure}

In order to discern excitonic effects on the
optical properties, we present first in 
Fig.~\ref{optics_ip} the optical spectra in the 
independent particle picture (IP)\cite{convergence}
(i.e. without the inclusion of the electron-hole coupling). This
yields an overview over relevant excitation mechanisms and later
allows one to trace down the effect of the electron-hole coupling.

We start the discussion with the case of $\text{C}_\text{N}$. Comparing the
corresponding panel of figure \ref{optics_ip} with the schematic
representation of the defect levels in the band gap
(Fig.~\ref{ldagw_levels}) one can identify the different absorption
peaks. The first band at $3.4$~eV is due to transitions form the
extended states at the valence edge to the defect levels within the
band gap; the other band at $4.4$~eV involves extended states below
the valence band edge and defect levels. Above $6$~eV we find the
bulk-like spectrum that is only slightly perturbed by the presence of
the defect. For the boron vacancy V$_{\text{B}}$ there are two bands
at $3.5$~eV and $4.7$~eV that are associated with transitions from
defect resonances in the valence band to the localized levels in the
band gap and a series of small peaks around $5.9$~eV originating from
valence band to defect transitions. 
The defect-related excitation bands of V$_\text{N}$ at $4.1$~eV, in contrast to the previous cases, involves extended states near the conduction band edge as final states. Only the faint defect-related excitation band around $6.6$~eV,
not visible in the figure, originates from valance band
states.
Another faint peak around $5$~eV is due to transitions among
the defect levels in the band gap. Finally, for the V$_\text{BN}$
di-vacancy, we find peaks between $5.6$~eV and $6.0$~eV due to
defect-defect transitions and valence-band to defect transitions,
respectively.

We now turn to the spectra\cite{convergence} including the electron-hole 
interaction via the Bethe-Salpeter equation. These are shown in
Fig.~\ref{optics_bse}.
The inclusion of the electron-hole
interaction has essentially two effects on the defect-related spectra: (i) a general
red-shift of the defect excitation energies due to electron-hole attraction and
(ii) a coupling of the independent particle excitations. This coupling
can give rise to bound electron-hole states, the excitons. In our
case the BSE leads to a red-shift of defect-defect transitions in the range
from $0.6$ to $1.3$~eV depending on the character and the localization
of the wavefunction. 
A common feature of all spectra is the large peak around 5.5\,eV which
corresponds to the strongly localized exciton of pure h-BN\cite{Arnaud2006,
PhysRevLett.96.126104,park2006} which is composed of
a large number of independent electron-hole pairs. In
the presence of a defect, electron-hole coupling leads to defect
bound excitons besides the ``free'' excitons. 

\begin{figure}[ht]
 \centering
 \includegraphics[width=0.27\textwidth,angle=-90]{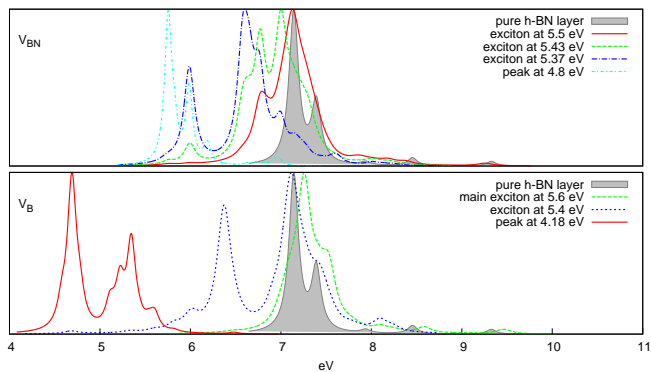}
 \caption{Amplitude of the different electron-hole pairing entering in
 the BSE, for the boron vacancy and the BN divacancy, \label{amp} compared with the pure h-BN monolayer.}
\end{figure}

\begin{figure*}[ht]
 \centering
 \includegraphics[width=0.68\textwidth,angle=-90]{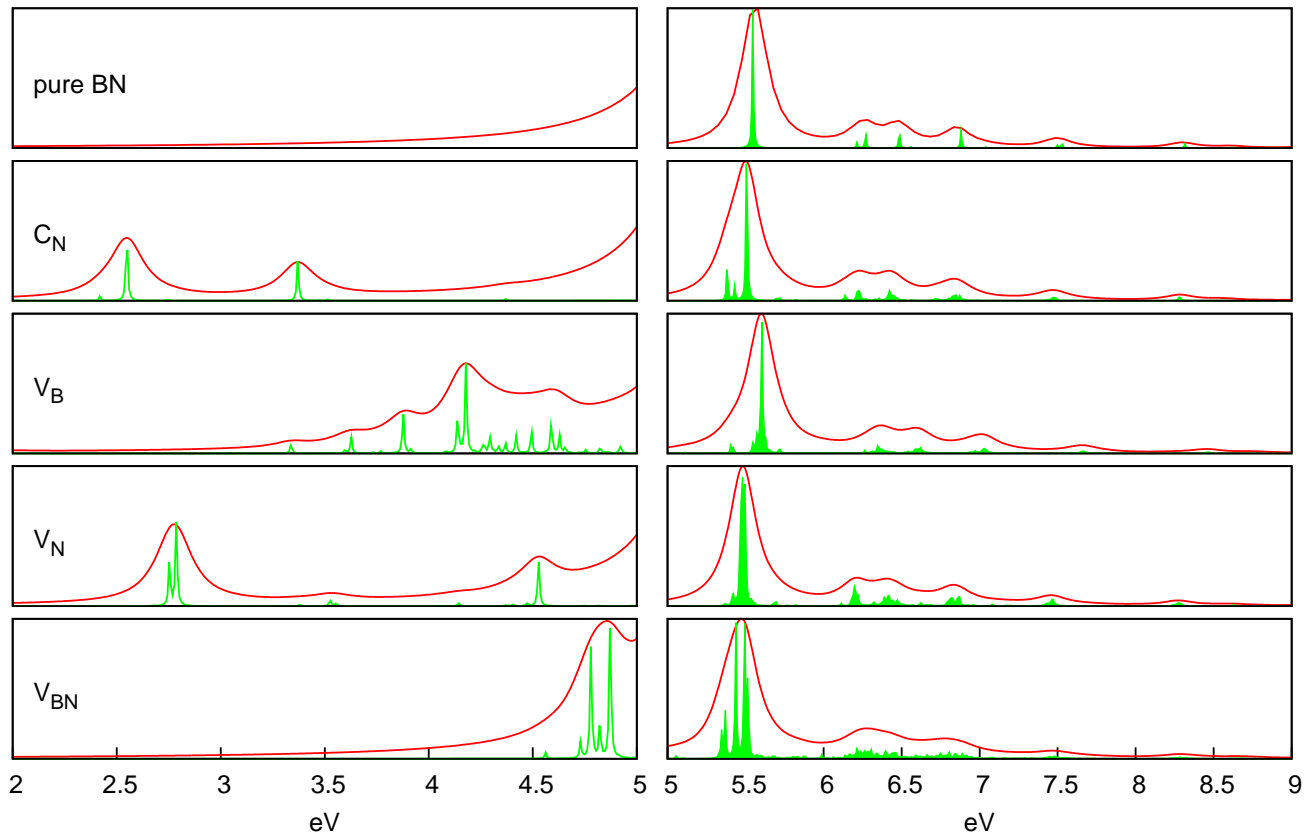}
 \caption{(color online) Optical absorption spectra calculated by the
 Bethe-Salpeter equation \label{optics_bse}. In the left panels
 the absorption cross section is increased by a factor of ten 
 in order to make the defect-related peaks visible. 
 Spectra are shown with a broadening of $0.1$~eV (to simulate a typical
 experimental broadening) and $0.005$~eV (in order to investigate the
 fine-structure of the spectrum).}
\end{figure*}

In the case of
$\text{C}_\text{N}$ and $\text{V}_\text{B}$, as one can see from
Fig.~\ref{optics_bse}, an additional small peak appears just below the
main excitons. In the case of $\text{V}_{\text{BN}}$ the main exciton
band has three sub-peaks due to the strong defect field with reduced
symmetry. In order to disentangle the nature of the different peaks,
we analyze the amplitude of the electron-hole pairs that compose each of
the exciton eigenstates $\lambda$, as a function of the energy
difference of the quasiparticle states of the corresponding electron
and hole states: \be A^\lambda(\omega) = \sum_{\eta = \{eh\}}
|\langle \eta | \lambda \rangle |^2 \delta(\omega - E_\eta), \ee
For brevity, we report $A^\lambda(\omega)$ only for
$\text{V}_{\text{B}}$ and $\text{V}_{\text{BN}}$ (see Fig.~\ref{amp}).  
The plots for the other defects show generally a similar behavior. 

Comparing for $\text{V}_{\text{B}}$ the electron-hole amplitudes of the four
excitation energies with the schematic diagram of
Fig.~\ref{ldagw_levels}, we can distinguish three kinds of excitations:
(i) transitions from valence bands or defect resonances to the defect
levels, that generate the peaks at $4.2$~eV, (ii) the main exciton at
$5.6$~eV that is entirely composed of extended states, and (iii)
a peak at $5.4$~eV generated by the coupling of bulk excitations 
with transitions from and to defect levels in the band gap.
The $\text{C}_{\text{N}}$ and $\text{V}_{\text{N}}$ cases present a structure similar to the one discussed
above.  From these results it is possible to infer three important
consequences: (i) transition from defect levels to valence/conduction
bands with an energy close to the band gap energy mixes with the bulk excitation
giving rise to an additional peak close to the main exciton; (ii)
transition between defect levels are renormalized by the
electron-hole interaction but do not mix with the bulk excitations;
(iii) the main exciton peak maintains a structure similar to the bulk
system.\\

The excitonic band of the di-vacancy $\text{V}_{\text{BN}}$ is
distinct from the other spectra as the di-vacancy possess both
acceptor and donor states of low symmetry in the vicinity of the band
edges. The direct transitions between acceptor and donor states give rise to
the absorption peaks around $4.8$~eV,
while the resonant coupling of transitions between the extended
conduction and valence band states with donor and acceptor states splits 
the main exciton in three sub-peaks respectively at $5.5$~eV, $5.43$~eV and $5.37$~eV (see Fig.~\ref{amp}).  

The observed side peaks to the main excitonic peak are compatible with 
the fine structure observed in different luminescence experiments\cite{PhysRevB.75.085205,jaffrennou:116102,Museur2008, PhysRevB.79.193104}.\\

\section{Conclusion}
\label{conclusion}
We have investigated the role of defects on the optical properties of BN
nanostructures. We have treated four different plausible defects in a
single sheet of hexagonal BN: a subsitutional carbon impurity,
a boron vacancy,
a nitrogen vacancy, and a BN di-vacancy.  Quasi-particle effects strongly modify
the defect-level positions within the band-gap and increase the exchange
splitting between occupied and unoccupied levels. Electron-hole interaction
(treated on the level of the the Bethe-Salpeter equation) not only leads to a
strongly bound excitonic state in pure BN but also strongly renormalizes
transitions from and to defect levels. This means that the peaks in the
optical spectra which are related to \textquotedblleft deep level impurities\textquotedblright 
are strongly affected by the e-e and e-h interaction and can be reliably calculated only on
the level of many-body perturbation theory (as opposed to the frequently used
random-phase approximation using DFT wave-functions and energies).
Experimental luminescence spectra (as presented, e.g., 
in Ref.~\onlinecite{jaffrennou:116102})
display a prominent peak close to $4$~eV with phonon replica on the low energy
side.  Comparing the four theoretical absorption spectra
(Fig.~\ref{optics_bse}) with the experimental luminescence spectrum, the best
agreement is obtained for the Boron vacancy where the main absorption peak of
the low energy regime is at 4.2 eV. The Stokes shift due to possible 
relaxation of the
excited state is not included in our calculation and limits the comparison of
experimental luminescence with theoretical absorption spectra.  In the case of
a strong (0.7 eV) Stokes shift, also the BN-divacancy (where a defect related
absorption peak is at 4.8 eV) could be responsible for the experimental
observation. The nitrogen vacancy (absorption peak at 4.5 eV) might also
explain the experiment. However, in this case a second strong luminescence
peak should also be observable below 3 eV which seems not to be the case in
experiments.

The 10$\times$10 supercell employed in our calculations is considerably
larger than the extension of the wave-function of the ``free'' exciton 
in pure h-BN.
Therefore, besides the defect-related absorption peaks, our calculations
also show the prominent absorption peak due to ``free'' excitions at 5.6 eV.
This peak is doubly degenerate in the pure BN sheet.
Coupling of the free exciton with one or several transitions from or to defect 
levels gives rise to a splitting of this peak and additional side peaks.
The details of the splitting depend on the nature of the
impurity but also on its periodic arrangement\cite{notedefstack}.
The fine-structure\cite{watanabe1} of the luminescence peak at 5.7 eV 
(with additional side peaks at 5.64 eV and 5.46 eV) can possibly be explained 
through our calculations which clearly display defect-related side peaks of 
the ``free'' excitonic peak. 
Furthermore, any kind of breaking of the exact hexagonal symmetry can lead 
to a splitting of the bright doubly degenerate ``free'' excitonic peak into 
two peaks and will also (for bulk h-BN) render the ``dark'' exciton 
(located 0.1 eV below
the bright exciton) visible \cite{PhysRevLett.100.189701}.
For reasons of computational feasibility, we have performed calculations 
only on point defects. Other kinds of defects such as dislocations and 
grain boundaries lead similarly to a breaking of symmetry and/or to the
formation of localized states and could thus as 
well explain the fine structure of luminescence spectra one would expect a
similar interaction of the ``free'' exciton with dangling-bond states). 
Indeed, Jaffrenou et 
al.\ had observed that the luminescence peaks at 5.64 eV and 5.46 eV occur 
mostly at grain boundaries and dislocations\cite{jaffrennou:116102} where 
defect states due to dangling bonds are supposed to occur.
The aim of our present study is thus not to unambiguously assign a certain
type of defect to the observed luminescence spectra, but rather to show the
general mechanism leading to the deep impurity levels and the formation
of defect bound excitons.
We also note that we did not study all possible point defects. A recent 
study of the chemical composition of an isolated h-BN sheet suggests the 
incorporation of oxygen atoms in the honeycomb lattice\cite{Krivanek}.
The same study also demonstrated that substitutional carbon atoms occur 
mostly in pairs, occupying neighboring nitrogen and boron sites.
Further experimental and theoretical studies are necessary to 
decide if the latter two types of defects can also occur during the synthesis
of BN nanostructures or if they are due to the experimental conditions
in the transmission electron microscope where vacancies in the h-BN sheet 
are induced by the electron bombardment 
and may be subsequently filled by a substitutional atom.

An alternative route to symmetry breaking was proposed by Watanabe et
al.\cite{PhysRevB.79.193104}. They suggested that the dynamical Jahn-Teller 
effect (splitting of the degenerate exciton by polaronic effects) could 
explain the excitonic splitting in hexagonal BN.
Further experimental and theoretical studies are necessary to disentangle 
the two effects.

\section*{Acknowledgments}
We acknowledge funding by the European Community through e-I3 ETSF
project (Contract Number 211956).  AR acknowledges funding by the
Spanish MEC (FIS2007-65702-C02-01), ACI-promciona project
(ACI2009-1036), "Grupos Consolidados UPV/EHU del Gobierno Vasco"
(IT-319-07). LW acknowledges funding by the French National Research 
Agency through project ANR-09-BLAN-0421-01.  
This work was performed using HPC resources from GENCI-IDRID (No. 100063 and  No. 091827). CA thanks X. Andrade and J. Alberdi Rodriguez for
the efficient computer-cluster installation in San Sebastian.
This work has been supported by the project no. FIS2010-21282-C02-01 (MCINN) from Spain.

\addcontentsline{toc}{chapter}{Bibliography}
\bibliographystyle{apsrev4-1}
\bibliography{defects}

\begin{thebibliography}{10}
\bibitem{DFT}%
  \BibitemOpen
  \bibinfo {note} {All DFT calculations have been performed using a
  10$\times$10 supercell containing $200$ atoms. The distance between the
  BN-sheet was 20 a.u. A single k-point was used. We used the
  exchange-correlation functional of Ref.~\onlinecite{ceperley}. For the
  calculations, we used the PWSCF code\cite{pwscf} with Troullier-Martins
  pseudopotetials \cite{troullier} and a plane wave cutoff of $50$~Ry. All the
  atomic structures considered in the paper were relaxed using the
  Broyden–Fletcher–Goldfarb–Shanno (BFGS) algorithm and a convergence
  threshold for the forces of $0.0001$~a.u. The 10$\times$10 k-point sampling
  of the elementary cell (which corresponds to the 1$\times$1 sampling of the
  10$\times$10 supercell) does not contain the high-symmetry point K, the point
  with the minimum direct gap. Therefore, the reported LDA gap of $4.82$~eV in
  Fig.~2 is slightly larger than the converged value
  $4.5$~eV\cite{wirtz_review}.}%
  \bibAnnoteFile{Stop}{DFT}%
\bibitem{GW}%
  \BibitemOpen
  \bibinfo {note} {Non-self consistent GW calculations have been performed
  starting from DFT-LDA wave-functions, using a plasmon pole approximation for
  the dielectric constant \cite{epsilon}, with the code YAMBO\cite{yambo}. We
  use a single k-point, 3000 bands for expanding the Green's function, 40.000
  G-vectors for the exchange and 35.000 for the correlation part of the
  self-energy. In contrast to Ref.~\onlinecite{PhysRevLett.96.126104} we did
  not add any scissor operator to correct the underestimation of the band gap
  in the $G_0W_0$ approximation.}%
  \bibAnnoteFile{Stop}{GW}%
\bibitem{epsilon}%
  \BibitemOpen
  \bibinfo {note} {The same dielectric constant has been employed in all the
  cases with and without defects, both for quasi-particle and Bethe-Salpeter
  equation. The dielectric constant has been calculated in the pure system,
  using 3000 bands, 3833 G-vectors for the dielectric matrix, and 35.000
  G-vectors for the wave-functions entering in the calculation.}%
  \bibAnnoteFile{Stop}{epsilon}%
\bibitem{BSE}%
  \BibitemOpen
  \bibinfo {note} {The optical spectra have been calculated using the
  Bethe-Salpeter equation (BSE) \cite{strinati}, with bands 300 to 500 (all
  $\pi$ and $\pi^*$ bands of the 10$\times$10 supercell), a $10$~eV cutoff on
  the electron-hole pair energy entering in the BSE and 20.000 G-vectors for
  the electronic wave-functions.}%
  \bibAnnoteFile{Stop}{BSE}%
\bibitem{convergence}%
  \BibitemOpen
  \bibinfo {note} {The independent-particle (IP) spectra are obtained for the
  1$\times$1 k-point set (corresponding to a 10$\times$10 sampling of the
  elementary unit cell) and are not fully converged with respect to k-point
  sampling for transitions involving extended band states, despite of the large
  supercells employed in the present calculations. Since we are only concerned
  with transitions involving localized defects levels, this is irrelevant for
  the analysis of excitation mechanism conducted here. Moreover we verified
  that peaks due to defects transition do not change increasing the number of
  k-points. In the BSE case, the main exciton collects most of the oscillator
  strengths. This speeds up the k-point convergence of the spectra and makes
  the presented spectra almost converged, contrary to the IP ones. In both
  cases IP and BSE we always considered the polarization parallel to the
  sheet.}%
  \bibAnnoteFile{Stop}{convergence}%
\bibitem{notedefstack}%
  \BibitemOpen
  \bibinfo {note} {In our simulation, we use a 60 degree unit cell which
  corresponds to a trigonal ordering of the defects. If a 90 degree super-cell
  was chosen, the ordering of the defects would be rectangular. Only in the
  limit of very large unit-cells, the splitting of the excitonic peak will be
  independent of the defect ordering and correspond to the coupling of a ``free
  excitons'' with an isolated defect. However, with the present computational
  ressources, we cannot reach the limit of isolated impurities by increasing
  the super-cell.}%
  \bibAnnoteFile{Stop}{notedefstack}%
\end{thebibliography}
\end{document}